\documentclass[preprint2,trackchanges]{aastex}

\usepackage{bbm}
\usepackage{amsmath}
\usepackage{cancel}
\DeclareMathOperator{\sgn}{sgn}

\begin{document}

\title{A Model for Dissipation of Solar Wind Magnetic Turbulence by Kinetic Alfv\'{e}n Waves at 
Electron Scales: Comparison with Observations}

\author{Anne Schreiner\altaffilmark{1} and Joachim Saur}
\affil{Institute of Geophysics and Meteorology\\ University of Cologne\\ 50923 Cologne, Germany}

\altaffiltext{1}{schreiner@geo.uni-koeln.de}

\begin{abstract}
In hydrodynamic turbulence, it is well established that the length of the dissipation scale depends 
on the energy cascade rate, i.e., the larger the energy input rate per unit mass, the more the turbulent 
fluctuations
need to be driven to increasingly smaller scales to dissipate the larger energy flux.
Observations of magnetic spectral energy densities indicate that this intuitive picture is not valid in solar wind turbulence.
Dissipation seems to set in at the same length scale for different solar wind conditions 
independently of the energy flux.
To investigate this difference in more detail,
we present an analytic dissipation model for solar wind turbulence at electron scales, 
which we compare with observed spectral densities. Our model combines the energy transport from large to small scales and collisionless damping, which 
removes energy from the magnetic fluctuations in the kinetic regime. We assume wave-particle interactions of 
kinetic Alfv\'{e}n waves (KAW) to be the main damping process. Wave frequencies and 
damping rates of KAW are obtained from the hot plasma dispersion relation.
Our model assumes a critically balanced turbulence, where larger energy cascade rates
excite larger parallel wavenumbers for a certain perpendicular wavenumber. If the dissipation 
is additionally wave driven such that the dissipation rate is proportional to the parallel 
wavenumber - as with KAW - then an increase of the energy cascade rate is counter-balanced by
an increased dissipation rate for the same perpendicular wavenumber leading to a dissipation length
independent of the energy cascade rate.
\end{abstract}


\section{Introduction} \label{sec:intro}
\label{sec_int}
Turbulence is a common feature in astrophysical and space plasmas, such as the interstellar medium, the solar wind or planetary magnetospheres. Turbulent processes are thought to play an
important role 
in cosmic ray propagation and energetic particle acceleration \citep[e.g.,][]{jok66,bie93,bie96,far04}. Furthermore turbulence and the 
associated dissipative processes could supply energy that is required to explain
the non-adiabatic temperature
profiles for the plasma species with increasing distance to the sun in the solar wind  \citep{ric95} 
and increasing distance to the central planet in the respective planetary magnetosphere
\citep{sau04,beg11,pap14}. 
The solar wind serves as a unique laboratory for 
in-situ measurements of space plasma turbulence thanks to numerous space missions \citep{bru13}. 
In the past decade, high time resolution magnetic field measurements taken by spacecrafts such as ACE, Cluster, or ARTEMIS 
led to a flurry of research activity to determine the characteristics of kinetic scale processes. 
But despite the
growing number of observed data sets,
there is still insufficient information to fully establish the properties of electron scale processes.
Additionally, due to the requirement for a kinetic description at these scales,
the interpretation of observations
with
the help of simulations and theoretical considerations remains particularly difficult.
Therefore a number of fundamental physical aspects of small scale solar wind turbulence
are still poorly understood.\\
It is well established that power spectra of magnetic fluctuations at magnetohydrodynamic (MHD) scales follow approximately the Kolmogorov scaling $k^{-5/3}$ 
\citep[e.g.,][]{mat82, den83, hor96, lea98, bal05}.
This spectral 
range is usually called the inertial range of solar wind turbulence. 
The first clear spectral break appears at ion scales, such as the ion Larmor radius or the ion inertial length \citep[e.g.,][]{lea99,ale08,che14}.
At these scales the physical mechanisms change leading
to a modification of the cascading process possibly including dissipation, which results in a modified spectral shape. At scales smaller than ion scales, a second cascade range up to electron scales with a steeper slope of about -2.9 to -2.3 is observed \citep{ale09, kia09, che10, sah10}, which is called the sub-ion range. Between the inertial range and the sub-ion range a transition region is observed, where the spectra exhibit a power law with a variable spectral index of -4 to -2 \citep{lea98,smi06,rob13} or a smooth non power law behavior \citep{bru14}. The steepening in the transition region has been associated with ion dissipation  \citep{smi12} or with the presence of coherent structures \citep{lio16}.
Even though Helios observations reached into the electron range \citep{den83}, it was only with the Cluster spacecraft
that the electron dissipation range was reached. So far there are only a few observations reported for such small scales with different interpretations \citep{ale09, sah09, sah10, ale12, sah13}.
A statistical 
study of magnetic power spectra by \citet{ale12} indicate an exponential spectral structure in the dissipation range and a universal behavior for all measured plasma parameters.
On the contrary, Cluster observations analyzed by \citet{sah13} indicate a third power law at the electron scales with a broad distribution of spectral indices varying from -5.5 to -3.5.
This result rather suggests 
a lack of universality of turbulent fluctuations in the dissipation range, however, the nature of the electron scale spectra and the associated universality remain an open issue.
\\
All these observations appear to be consistent with an important role of kinetic Alfv\'{e}n waves (KAW). 
The following picture of a KAW generated turbulent cascade is presented in the literature:
In the inertial range nonlinear interactions between Alfv\'{e}n waves are responsible for the
generation of the turbulent cascade. 
At scales comparable to the ion Larmor radius, the Alfv\'{e}n wave is possibly slightly damped, which would explain the transition range \citep{den83}.
However, the process that leads to a steepening of the spectrum in the sub-ion range, i.e., between ion and electron scales  is the transformation from the  non-dispersive Alfv\'{e}n wave to the dispersive KAW \citep{how06}. 
The energy in Alfv\'{e}nic fluctuations generates a dispersive KAW cascade down to 
the electron scales, which again can be described in fluid-like terms \citep{sch09}.
In the vicinity of the electron Larmor radius or the electron inertial length, the KAW is subject to strong Landau damping via wave-particle interactions \citep{gar04,sah09}.
Since properties of the whistler wave are similar to those of the KAW \citep[e.g.,][]{bol13}, it is difficult to distinguish these waves in observations. Hence, there is still 
an ongoing debate whether the
small scale fluctuations consist of whistler waves or KAW \citep{gar09, sal12, che13}.\\
Observations with different angles between the mean magnetic field and the solar wind flow direction lead to the understanding that magnetic fluctuations are anisotropic with respect to the 
mean magnetic field in both the MHD regime \citep[e.g.,][]{bar79, mat90, bie96, hor08}
and the kinetic regime \citep[e.g.,][]{che10, sah10, nar11}. \citet{gol95} proposed a particular model for the anisotropy, called critical
balance, which leads to observed $P(k_{\perp})\propto k_{\perp}^{-5/3}$ and 
$P(k_{\parallel})\propto k_{\parallel}^{-2}$ spectra in the inertial range \citep{hor08, pod09}. By equating the nonlinear timescale at which the energy is transferred to smaller 
scales with the linear Alfv\'{e}n timescale, one finds
$k_{\parallel}\propto k_{\perp}^{2/3}$ in the inertial range and $k_{\parallel}\propto k_{\perp}^{1/3}$ in the kinetic range \citep{cho04, sch09}. Hence, the turbulence
becomes more anisotropic for high wavenumbers
and the energy is cascaded mainly in the perpendicular direction $k_{\perp}\gg k_{\parallel}$. Although recent observations and simulations are 
consistent with the critical balance assumption \citep{ten12,he13,pap15}, its applicability 
to solar wind turbulence is still subject of debate 
and other models are proposed to explain the anisotropy \citep{nar10, li11, hor12, wan14, nar15a}.\\
A surprising result in the observation by \citet{ale12} is the independence
of the dissipation length from the amplitude
of the turbulent spectra $P_0$ at a fixed wavenumber $k_0$. 
This independence is a remarkable difference compared to
hydrodynamic turbulence, where the dissipation length $l_{d,\mathrm{Kolm}}= (\nu^3/\varepsilon_0)^{1/4}$ is given by the 
energy cascade rate $\varepsilon_0$ and the kinematic viscosity
$\nu$ \citep[e.g.,][]{fri95}.
Accordingly, in hydrodynamic turbulence, the more energy is injected per unit mass, the 
more the turbulence is driven to smaller scales to dissipate the
larger energy flux.
Following \citet{kol41}, the amplitude of the turbulent spectra and the
energy cascade rate are related by $\varepsilon_0\propto P_0^{3/2} k_0^{5/2}$.
The solar wind observations by \citet{ale12} show approximately no dependence 
of the dissipation length
on the energy cascade rate. This is indeed surprising under the assumption 
that the energy is not fully dissipated 
at a resonance, but that the dissipation rate $\gamma$ is a smooth
function of wavenumber $k$ such as, e.g., for
Landau damping of KAW \citep{lys96, how06, sah12, nar15}. In this case,
one would still expect that a larger energy flux drives the turbulence to smaller scales
before the energy is dissipated.
This effect is neither noted nor discussed
in earlier dissipation models by \citet{how08, pod10, how11}, although
the independence of the dissipation length scale from the energy cascade rate
is implicitly included in these models.
To discuss this issue in detail, we present a 'quasi' analytical dissipation
model to describe magnetic power spectra at sub-ion scales.
The model 
is tailored to be applied for data comparison with variable spectral 
slope and associated critical balance. 
For the description of the turbulent energy transport, we introduce in Section 
\ref{sec_mod} a cascade model, which is in several aspects 
similar to earlier turbulence models \citep[e.g.,][]{pao65, how08, pod10, zha13}.
Still, we give a short derivation of our model equation in order to establish a 
basis for theoretical predictions of solar wind dissipation processes and to discuss
the independence of the dissipation length from the energy cascade rate.
As a damping rate, we include the imaginary part of the KAW wave frequency obtained from linear Vlasov theory.
In Section \ref{sec_imp}, we investigate the dissipation length scale and the spectral
shape of the dissipation range under the assumption 
of linear KAW damping and critically balanced turbulence. In Section \ref{sec_app}, we present a statistical study, where we fit
an exponential function proposed by \citet{ale12} to 300 model spectra for
varying solar wind conditions. In Section \ref{sec_dis}, we discuss the limitations of 
our approach and of the resultant implications for solar wind dissipation.

\section{MODEL FOR MAGNETIC ENERGY SPECTRA}
\label{sec_mod}
In this section, we construct a dissipation model for energy spectra of turbulent fluctuations. The model is a linear combination of 
the nonlinear transport of energy from the 
large to the small scales and the dissipation process at small scales. In its general form, the model can in principle describe 
turbulent spectra in any plasma or fluid.
For solar wind turbulence we assume a critically balanced energy cascade of KAWs up to the highest wavenumbers 
where the energy is dissipated by wave-particle interactions. Turbulent dissipation is quantified by the imaginary part of the wave 
frequency obtained from 
a dispersion relation for KAWs. Note that similar to common terminology in previous publications, the term "dissipation" refers in this paper to the transfer of energy from the magnetic field into perturbations of the particle distribution function via wave-particle interactions. The final transfer of this non-thermal free energy in the distribution function to thermal energy, i.e, the irreversible thermodynamic heating of the plasma, can only be achieved by collisions
\citep{sch09,how15, sch16}. 
\subsection{Energy Cascade and Dissipation}
\label{sec_modequ}
Based on the idea that the turbulent energy cascades self-similarly to higher wavenumbers \citep{fri95}, we write the energy cascade rate as
\begin{linenomath*}
\begin{equation}
 \varepsilon(k) = C_K^{-3/2} P(k) v_k(k),
 \label{equ_neu1}
\end{equation}
\end{linenomath*}
where $P(k)$ defines the spectral energy density of magnetic fluctuations and $C_K$ is the dimensionless Kolmogorov constant. 
We introduce the 'velocity' of the energy transport in wavenumber space or 'eddy-decay velocity' $v_k(k) = dk/dt$.
In the inertial range, the energy cascade rate $\varepsilon_0$ is constant, i.e., the energy is transported loss-free from large to small scales.
In this case, (\ref{equ_neu1}) can be written as
\begin{linenomath*}
\begin{equation}
 \varepsilon_0 = C_K^{-3/2} P_0 v_{k0} = \mbox{const.,} \label{equ1}
\end{equation}
\end{linenomath*}
where $P_0 = P(k_0)$ and $v_{k0} = v_k(k_0)$
characterize the spectral properties at a 
wavenumber $k_0$ in the inertial range.
The fluid velocity $v$ and the eddy-decay velocity of magnetic fluctuations $v_k$ are related by 
\begin{linenomath*}
\begin{equation}
v_k(k) = \frac{dk}{dt}= k^2 v(k).
 \label{equ10}
\end{equation}
\end{linenomath*}
The ratio of velocity to magnetic fluctuations $\alpha$ is assumed to be \citep{sch09}
\begin{linenomath*}
\begin{equation}
v(k) = \alpha \sqrt{\frac{P(k)k}{\rho}},
\label{equ_alpha}
\end{equation}
\end{linenomath*}
with the mass density $\rho$.
From (\ref{equ_neu1}), (\ref{equ10}), and (\ref{equ_alpha}), we obtain
\begin{linenomath*}
\begin{equation}
 P(k) = C_K \rho^{1/3} \varepsilon(k)^{2/3} \alpha(k)^{-2/3} k^{-5/3}.
 \label{equ_P}
\end{equation}
\end{linenomath*}
Assuming $\alpha$ to follow a power law of the form $\alpha= \alpha_0 (k/k_0)^{\beta}$, we can write $P(k)$ as
\begin{linenomath*}
\begin{equation}
P(k) = P_0 \left(\frac{\varepsilon(k)}{\varepsilon_0}\right)^{2/3} \left(\frac{k}{k_0}\right)^{-\kappa},
 \label{equ_P2}
\end{equation}
\end{linenomath*}
with $\kappa=2/3\beta+5/3$.
With (\ref{equ_neu1}), (\ref{equ1}), and (\ref{equ_P2}), we write the eddy-decay velocity $v_k(k)$ as:
\begin{linenomath*}
\begin{equation}
 v_k(k) = v_{k0} \left(\frac{\varepsilon(k)}{\varepsilon_0}\right)^{1/3}\left(\frac{k}{k_0}\right)^{\kappa}.
 \label{equ2}
\end{equation}
\end{linenomath*}
Due to dissipation, the energy flux at wavenumber $k'=k+dk$ differs from the energy flux at
$k$ by the part of energy $D(k)dk$ that is dissipated 
\begin{linenomath*}
\begin{equation}
C_K^{-3/2} P(k) v_k(k) = C_K^{-3/2} P(k') v_k(k')+D(k)dk.\label{equ3}
\end{equation}
\end{linenomath*}
The heating rate $D(k) = 2 P(k) \gamma(k)$ contains a damping rate $\gamma(k)$.
From (\ref{equ_P2}), 
(\ref{equ2}), (\ref{equ3}), and a Taylor expansion of $P(k') v_k(k')$ for small  $dk$ in equation (\ref{equ3}), 
we obtain a differential equation for the energy spectrum of turbulent fluctuations $P(k)$
\begin{linenomath*}
\begin{equation}
 \frac{d P(k)}{d k} = -P(k) \left( \frac{\kappa}{k} + \frac{4}{3} C_K^{3/2} \frac{\gamma(k)}{v_k(k)} \right).
 \label{equ4}
\end{equation}
\end{linenomath*}
The solution of (\ref{equ4}) for $P(k)$ yields the one-dimensional energy spectrum
\begin{linenomath*}
\begin{align}
 P(k) =& P_0 \left(\frac{k}{k_0}\right)^{-\kappa} \exp \left(-\frac{4}{3} C_K^{3/2} \right. \nonumber  \\   & \times \left. \int_{k_0}^{k} dk'\frac{\gamma(k')}{v_k(k')}  \right). \label{equ5}
\end{align}
\end{linenomath*}
Insertion of (\ref{equ_neu1}) and (\ref{equ_P}) in (\ref{equ5}) leads to
\begin{linenomath*}
\begin{align}
 P(k) =& P_0 \left(\frac{k}{k_0}\right)^{-\kappa} \exp \left(-\frac{4}{3} C_K \int_{k_0}^{k} dk'
 \left(\frac{\varepsilon(k')}{\rho}\right)^{-1/3} \right. \nonumber \\ &
 \left. \times\, \alpha(k')^{-2/3}\gamma(k') k'^{-5/3}\right). \label{equ5b}
\end{align}
\end{linenomath*}
With (\ref{equ_P2}), equation (\ref{equ5b}) can be written in terms of the energy flux
\begin{linenomath*}
\begin{align}
 \varepsilon(k) =& \varepsilon_0 \exp \left(-2 C_K \int_{k_0}^{k} dk'
 \left(\frac{\varepsilon(k')}{\rho}\right)^{-1/3} \right. \nonumber \\ 
& \times\, \left. \alpha(k')^{-2/3}\gamma(k') k'^{-5/3}\right).
 \label{equ5c}
\end{align}
\end{linenomath*}
Under the assumption that the eddy-decay velocity is not affected by the dissipation
\begin{linenomath*}
\begin{equation}
 v_k(k) \approx v_{k0} \left(\frac{k}{k_0}\right)^{\kappa},
 \label{equ_neu6}
\end{equation}
\end{linenomath*}
and
using (\ref{equ1}) and (\ref{equ_P}), equation (\ref{equ5}) simplifies to
\begin{linenomath*}
\begin{align}
P(k) =& P_0 \left(\frac{k}{k_0}\right)^{-\kappa}
\exp \left(-\frac{4}{3} C_K \left(\frac{\varepsilon_0}{\rho}\right)^{-1/3} \alpha_0^{-2/3} \right. \nonumber \\  & \times\, \left. k_0^{-5/3}\int_{k_0}^{k} dk' \gamma(k') \left(\frac{k'}{k_0}\right)^{-\kappa}\right).
\label{equ_P3}
 \end{align}
 \end{linenomath*}
Turning to hydrodynamic turbulence and insertion of a resistive damping rate $\gamma(k) = \nu k^2$ with the kinematic viscosity $\nu$, which is valid in a collisional fluid \citep[e.g.,][]{dra06}, we can use our model to calculate the associated energy spectrum. When we assume that the 
eddy-decay velocity is not affected by the damping as in (\ref{equ_neu6}), we find 
\begin{linenomath*}
\begin{equation}
 P(k) = P_0 \left(\frac{k}{k_0}\right)^{-5/3} \exp\left(- C_K \nu \left(\frac{\varepsilon_0}{\rho}\right)^{-1/3} k^{4/3}\right),
 \label{equ6}
\end{equation}
\end{linenomath*}
where we use $\kappa = 5/3$, $k_0 \ll k$, $\alpha_0 =1$, and where $P(k)$ denotes the energy
density of velocity fluctuations in this case.
This spectral form has been found previously by \citet{cor64} and \citet{pao65}. Equating the length scale, 
where the argument of the exponential function in equation (\ref{equ6}) assumes -1, we obtain the 
dissipation scale for hydrodynamic turbulence
\begin{linenomath*}
\begin{equation}
 l_{d,hd} = C_K^{3/4} \left(\frac{\nu^3 \rho}{\varepsilon_0} \right)^{1/4},
 \label{equ7}
\end{equation}
\end{linenomath*}
which is apart from constant factors on the order of unity in agreement
with the Kolmogorov dissipation scale $l_{d,\mathrm{Kolm}}\sim (\nu^3/\varepsilon_0^*)^{1/4}$ with the cascade rate per unit mass
$\varepsilon_0^*= \varepsilon_0/\rho$. 
Assuming alternatively that the eddy-decay velocity is slowed down by the damping in the dissipation range according to (\ref{equ2}), we find an algebraic spectral energy density
\begin{linenomath*}
\begin{equation}
P(k)=P_0 \left(\frac{k}{k_0}\right)^{-5/3} \left(1-\frac{1}{2}C_K \left(l_{d,hd} k\right)^{4/3}\right)^2,
\label{equ_HD}
\end{equation}
\end{linenomath*}
where we again use $\kappa=5/3$, $k_0\ll k$, and $\alpha_0=1$. $P(k)$ decreases more rapidly compared to the previous case and vanishes at a maximum wavenumber. A similar spectral form has been found by \citet{kov48}. Expressions (\ref{equ6}) and (\ref{equ_HD}) provide models how the dissipation and the associated dissipation length depend on the energy flux in hydrodynamic turbulence.
Consequences resulting from this fact and differences to solar wind
turbulence will be discussed in 
Section
\ref{sec_imp}.
\\
For KAWs, we include the normalized damping rate
\begin{linenomath*}
\begin{equation} \gamma(k_{\perp}, k_{\parallel})= k_{\parallel} v_A\overline{\gamma}(k_{\perp}, k_{\parallel}) ,
\label{equ8}
\end{equation}
\end{linenomath*}
which is the imaginary part of the complex wave frequency
in the dispersion relation for KAWs with $\omega=\omega_r+\mathrm{i}\gamma$ and the Alfv\'{e}n velocity $v_A = B_0/ \sqrt{\mu_0 \rho}$. 
We assume that the linear Alfv\'{e}n time scale and the nonlinear time scale are equal at all scales.
This equality is the critical balance assumption of \citet{gol95}, which leads to a relation between $k_{\parallel}$ and $k_{\perp}$
\begin{linenomath*}
\begin{equation}
 v_{\perp}(k_{\perp}) k_{\perp} = k_{\parallel} v_{ph,A} =
 k_{\parallel} v_A \overline{\omega}_r,\label{equ9}
\end{equation}
\end{linenomath*}
where $v_{\perp}$ is the plasma velocity perpendicular to the mean magnetic field, which we take in the remainder as the turbulent velocity 
fluctuations introduced in (\ref{equ10}) and (\ref{equ_alpha}), $v_{ph,A} = v_A \overline{\omega}_r$ 
is the phase velocity of the wave, and $\overline{\omega}_r
=\omega_r/k_{\parallel}v_A$ is the real
part of the normalized wave frequency describing the deviations from the MHD shear Alfv\'{e}n wave. 
From (\ref{equ_neu1}), (\ref{equ10}), (\ref{equ_P}), and (\ref{equ9}),
we obtain an equation for the parallel wavenumber as a function of the perpendicular wavenumber
\begin{linenomath*}
\begin{equation}
 k_{\parallel} =  C_K^{1/2}(v_A \overline{\omega}_r)^{-1} \left(\frac{\varepsilon(k_{\perp})}{\rho}\right)^{1/3} 
 \alpha(k_{\perp})^{2/3} k_{\perp}^{2/3}.
 \label{k_parallel}
\end{equation}
\end{linenomath*}
For $\alpha(k_{\perp}) \approx \overline{\omega}_r$ \citep{how08} and without dissipation 
($\varepsilon(k_{\perp})=\varepsilon_0$), (\ref{k_parallel}) leads
to the typical relations for $k_{\parallel}$ and $k_{\perp}$ as discussed in the
introduction in both the MHD regime ($\overline{\omega}_r \approx 1$) and the
kinetic regime ($\overline{\omega}_r \approx k_{\perp}\rho_i$).
Inclusion of (\ref{equ8}) and (\ref{k_parallel}) into (\ref{equ5b}) yields the perpendicular energy spectrum for magnetic fluctuations
\begin{linenomath*}
\begin{align}
P(k_{\perp}) =& P_0 \left(\frac{k_{\perp}}{k_0}\right)^{-\kappa} \exp \left( -\frac{4}{3} C_K^{3/2} \right. \nonumber \\
& \times\, \left. \int_{k_0}^{k_{\perp}} dk_{\perp}' \frac{\overline{\gamma}(k_{\perp}',k_{\parallel})}{\overline{\omega}_r(k_{\perp}',k_{\parallel})} k_{\perp}'^{-1} \right).
\label{equ12}
\end{align}
\end{linenomath*}
Again with (\ref{equ_P2}), equation (\ref{equ12}) can be expressed in terms of the energy flux
\begin{linenomath*}
\begin{equation}
 \varepsilon(k_{\perp}) = \varepsilon_0 \exp \left( -2 C_K^{3/2}
\int_{k_0}^{k_{\perp}} dk_{\perp}' \frac{\overline{\gamma}(k_{\perp}',k_{\parallel})}{\overline{\omega}_r(k_{\perp}',k_{\parallel})} k_{\perp}'^{-1} \right).
\label{E_SW}
\end{equation}
\end{linenomath*}
The latter expression is apart from constant factors
similar to the dissipation model proposed by \citet{how08}. From (\ref{equ10}), (\ref{equ5}),
(\ref{equ8}), and (\ref{equ9}), we see that the energy spectrum in (\ref{equ12}) and the associated energy flux in (\ref{E_SW}) are independent of the choice of the eddy-decay velocity in the dissipation range, i.e., it leads to the same results for (\ref{equ2}) and (\ref{equ_neu6}).
\subsection{Damping Rates of Kinetic Alfv\'{e}n Waves} \label{sec_damp}
In this section, we present the calculation of damping rates obtained from the hot plasma dispersion relation for 
a nonrelativistic plasma with Maxwellian distributed electrons and protons with no zero-order drift velocities.
The hot plasma dispersion relation
refers to the general relationship arising from the set of linearized Vlasov Maxwell equations \citep[e.g.,][]{sti92}
\begin{linenomath*}
\begin{equation}
\mathrm{det} \left[ {\bf k} \otimes  {\bf k} - k^2 \mathbbm{1} + \frac{\omega^2}{c^2} \underline{\underline{\epsilon}} \right]=0,
\end{equation}
\end{linenomath*}
where $\mathbbm{1}$ denotes the identity matrix, $c$ the speed of light, and $\epsilon_{ij}$ the 
elements of the dielectric tensor (see Appendix \ref{app_tensor} for
a description of the dielectric
tensor elements and definitions of all symbols).
Assuming that the wave vector is in the $xz$ plane, the dispersion relation can be written in the form
\begin{linenomath*}
\begin{eqnarray}
  \mathrm{det} \left(\begin{array}{ccc}
                       \epsilon_{xx}-n_{\parallel}^2 & \epsilon_{xy} & \epsilon_{xz}+n_{\parallel}n_{\perp}\\
                       -\epsilon_{xy}                & \epsilon_{yy} -n^2 &    \epsilon_{yz}\\
                       \epsilon_{xz}+n_{\parallel} n_{\perp}  & -\epsilon_{yz} & \epsilon_{zz} - n_{\perp}^2
                      \end{array}\right)
   = 0,
  \label{equ_det}
 \end{eqnarray}
 \end{linenomath*}
with the parallel, perpendicular and total index of refraction $n_{\parallel}=k_{\parallel} c /\omega$, $n_{\perp}= k_{\perp} c /\omega$ and 
$n= k \omega/c$, respectively.
From equation (\ref{equ_det}), we obtain the wave frequency as a complex number, 
$\omega=\omega_r + \mathrm{i} \gamma$. Details of the numerical evaluation are given
in Appendix \ref{app_num}.\\
We compare the resultant damping rates with 
two other damping rates for KAW: Damping rates obtained from the hot dispersion relation
with the Pad\'{e}
approximation for the plasma dispersion function $Z(\xi)$, which is used in other dispersion relation solvers \citep[e.g.,][]{rön82, nar15},
and damping rates from a simplified algebraic dispersion relation found by \citet{lys96}, which was derived to describe low-frequency waves in small plasma beta plasmas, e.g., the Earth's magnetosphere. 
The advantage of both methods are much faster computation times of the root finding algorithm in comparison to the hot dispersion relation solver.
For low-frequency waves ($\omega\ll \Omega_s=q_s B/m_s$, with gyrofrequency $\Omega_s$, particle charge $q_s$ and particle mass $m_s$ for species $s$), large parallel wavelength 
($k_{\parallel}v_s \ll \Omega_s$, with thermal velocity $v_s$), and small plasma betas ($\beta_s =2 k_B T_s n_s \mu_0 /B^2\ll1$, with temperature $T_s$ and  number density $n_s$)
the full system of the hot dispersion relation can be approximated by a $2\times2$ matrix since the fast mode can be factored out.
Then the determinant of the 
$2\times2$ matrix yields the dispersion relation \citep{lys96}
\begin{linenomath*}
\begin{equation}
\frac{\omega^2}{k_{\parallel}^2v_A^2} = \frac{k_{\perp}^2\rho_i^2}{1-\Gamma_0(k_{\perp}^2\rho_i^2)}+\frac{k_{\perp}^2\rho_a^2}{\Gamma_0(k_{\perp}^2 \rho_e^2)\left[1+\xi Z(\xi)\right]},
\label{equ16}
\end{equation}
\end{linenomath*}
with the gyroradius $\rho_s =v_s/\Omega_s$ and the ion acoustic gyroradius $\rho_a^2= k_B T_e/m_i\Omega_i^2$ (see Appendix \ref{app_tensor} for definitions of all other symbols).
Note that $\xi = \xi(\omega)$; thus, equation (\ref{equ16}) is an implicit
equation for the normalized wave frequency $\overline{\omega} = \overline{\omega}_r + \mbox{i} \overline{\gamma}$,
which can be solved numerically.
Figure \ref{fig_LL96T} shows normalized damping rates ($\overline{\gamma}/\overline{\omega}_r$) calculated
from the hot dispersion relation (solid lines), the hot
dispersion relation with Pad\'{e} 
approximation (dotted lines), and the \citet{lys96} approximation (dashed lines) for temperature ratios 
of $T_i/T_e=1$ (panel (a)) and 
$T_i/T_e=10$ (panel (b)) for ion plasma beta values of 0.01, 0.1, 1, and 10.
The ratio of $k_{\parallel}$ to $k_{\perp}$ is given
through the critical balance condition in (\ref{k_parallel}).
We use typical solar wind values for the magnetic field (10 nT)
and the electron number density 
(10 cm$^{-3}$). 
\begin{linenomath*}
\begin{figure}
 \gridline{\fig{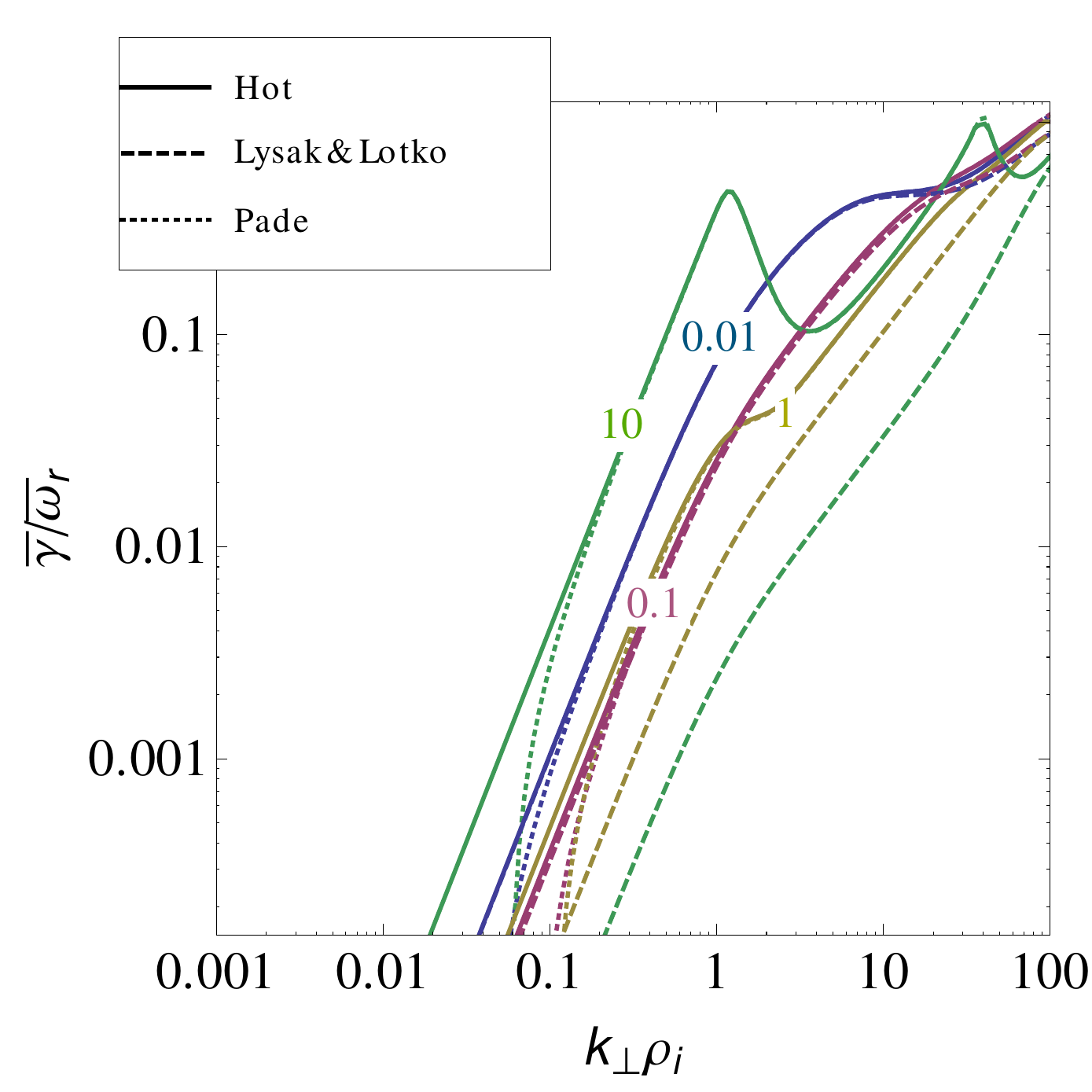}{0.5\textwidth}{(a)}}
\gridline{\fig{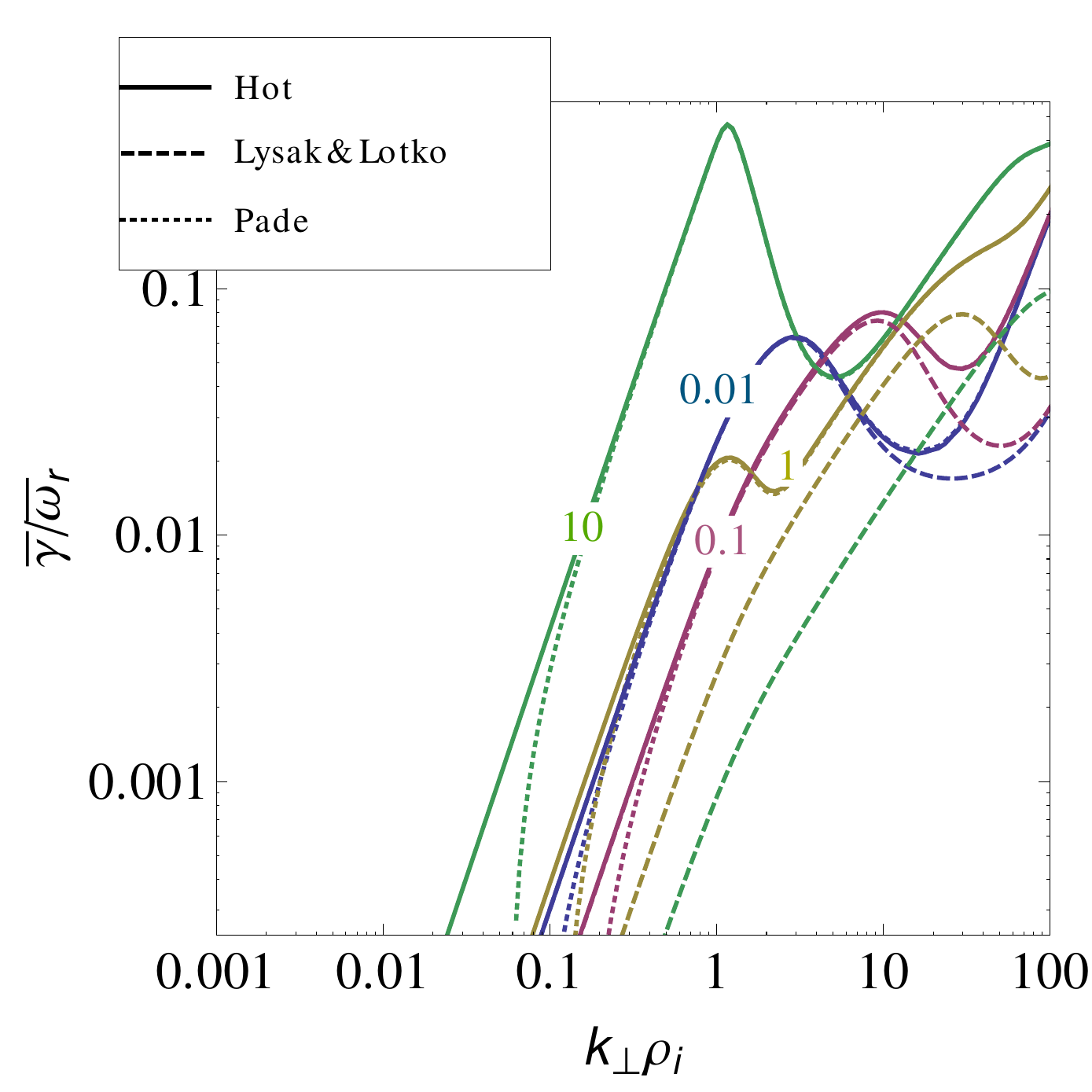}{0.5\textwidth}{(b)}}
  \caption{$\overline{\gamma}/\overline{\omega}_r$ for a temperature ratio of $T_i/T_e=1$ (a) and of
  $T_i/T_e=10$ (b) and ion plasma betas of 0.01, 0.1, 1.0 and 10.0. Damping rates from the simplified dispersion relation are shown in dashed lines, 
  hot damping rates in solid lines and hot damping rates with Pad\'{e} approximation in dotted lines.}
  \label{fig_LL96T}
 \end{figure}
 \end{linenomath*}
For all values of $\beta_i$, hot damping rates with Pad\'{e} approximation are in agreement with hot damping rates for $k_{\perp}\rho_i>1$,
but show small errors when the wave frequency is almost real and $\gamma$ is nearly negligibly small.
Due to critical balance, the real part of the wave frequency does not reach the ion gyrofrequency where 
differences of the plasma dispersion function and the Pad\'{e} approximation would occur.
Damping rates calculated with the \citet{lys96} approximation show good agreement with hot damping rates for $\beta_i=0.01$ and $\beta_i=0.1$. 
Small deviations occur at $k_{\perp}\rho_i\approx 10$, 
where $\omega_r$ comes closer to the ion gyrofrequency.
For $\beta_i\ge 1$, both the amplitude and the general form of the damping rates 
calculated with the \citet{lys96} approximation differ significantly from hot damping rates already for scales $k_{\perp}\rho_i<1$.
The results confirm that the \citet{lys96} dispersion relation can be well applied for $\beta_i\ll1,~\beta_e\ll1$ and $\omega_r\ll\Omega_i$.
Although the simplified dispersion relation is valid for a range of solar wind parameters,
quantitative conclusions concerning damping at electron scales 
cannot be drawn. For a complete analysis of dissipation processes under the full parameter space of the solar wind conditions usage of the hot dispersion relation is necessary.
\subsection{Implications for the Dissipation Range}
\label{sec_imp}
\begin{figure}[ht!]
 \plotone{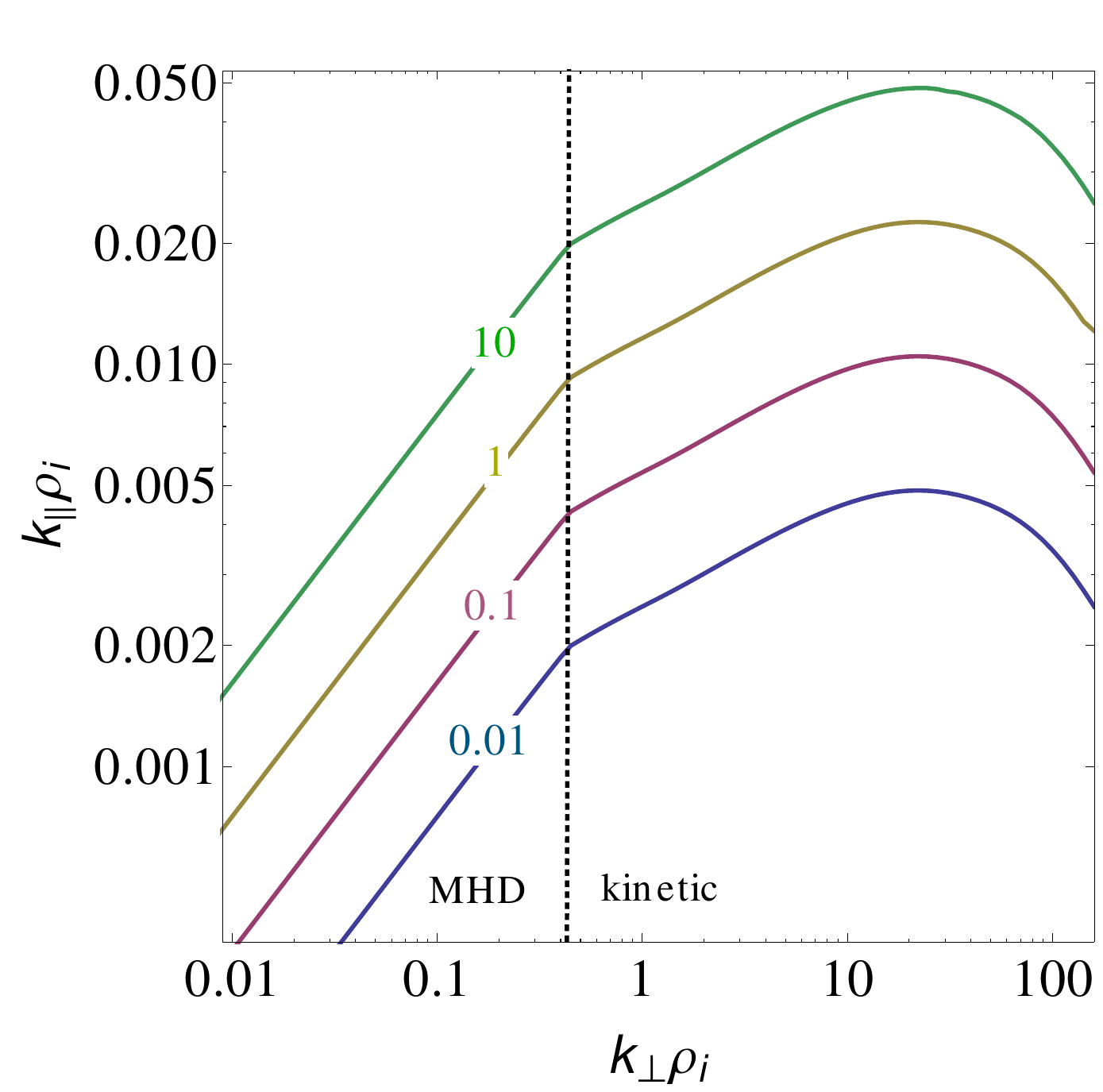}
 \caption{Equation (\ref{k_parallel}) for
 $\varepsilon_0= \lbrace 0.01, 0.1, 1, 10\rbrace \times \varepsilon_{0,\mathrm{ref}}$ 
 with $\varepsilon_{0,\mathrm{ref}}= 7\times10^{-16} \ \rm J m^{-3} s^{-1}$ calculated from 
\citet{ale09}.
 The dotted line shows the transition from MHD to the kinetic regime. Solar wind parameters 
 ($B=15.5$ nT, $n=20$ cm$^{-3}$, $T_i = 61$ eV, $T_e=26$ eV, and $v_S = 630$ km/s), the 
break frequency, and $\varepsilon_{0,\mathrm{ref}}$ are taken from observation 5 in \citet{ale09}.}
\label{fig_kpara}
\end{figure}

\begin{figure}[ht!]
 \plotone{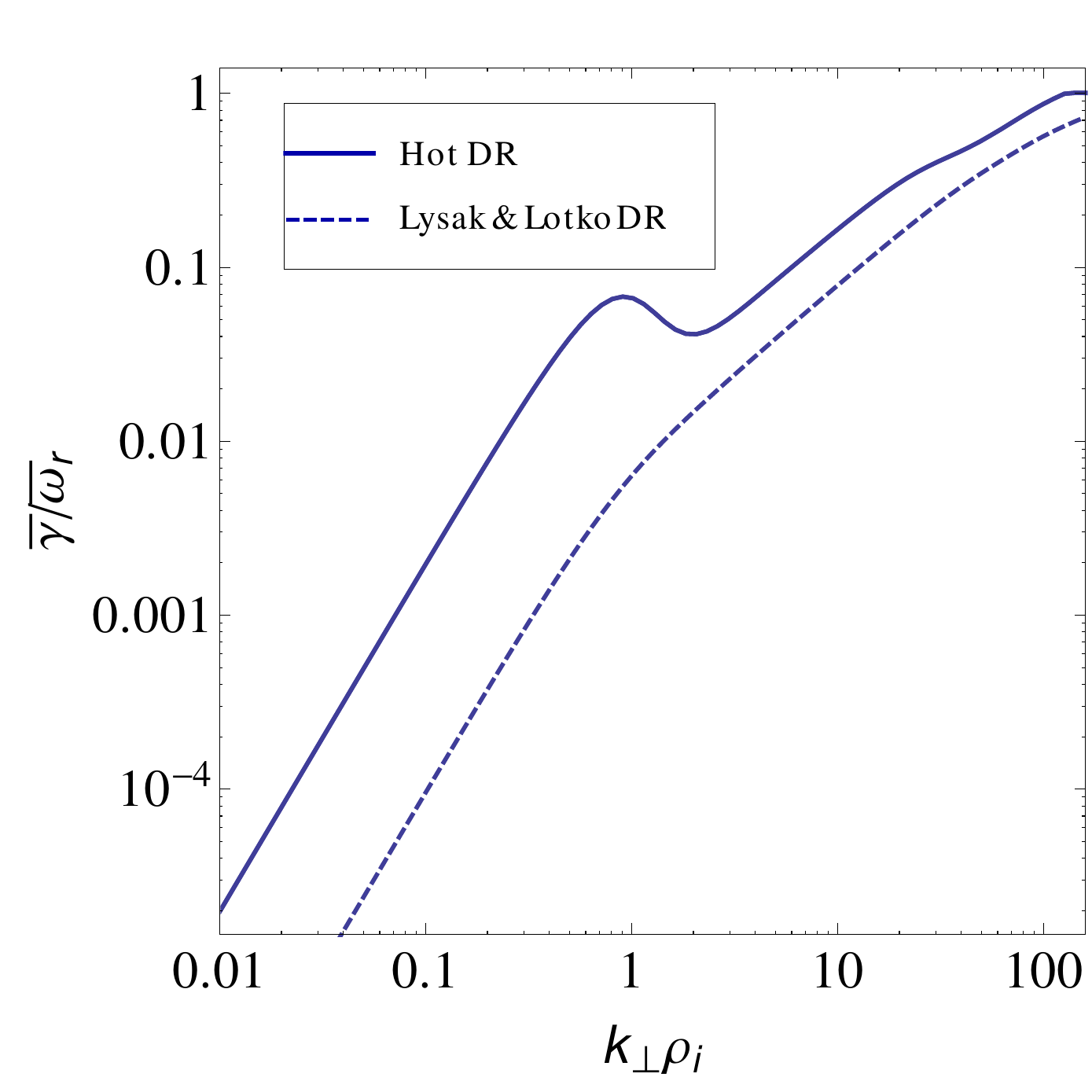}
 \caption{Hot damping rate ($\overline{\gamma}/\overline{\omega}_r$) for all 
ratios of $k_{\parallel}$ to $k_{\perp}$ from Figure \ref{fig_kpara} and
the same parameters as in Figure \ref{fig_kpara}. All damping rates fall approximately on the same dark blue solid line. 
$\overline{\gamma}/\overline{\omega}_r$ from the \citet{lys96} approximation is shown
in the dashed line for comparison.}
\label{fig_gamma_epsilon}
\end{figure}

With our model for the spectral energy density in equation (\ref{equ12}) we 
can draw conclusions about the dissipation length and the spectral shape 
of the solar wind dissipation range.
Let us first look at the critical balance assumption in (\ref{k_parallel}) again.
Equation
(\ref{k_parallel}) reveals the dependence of
the parallel wavenumber on the energy flux $\varepsilon(k_{\perp})$.
Consequently, $\gamma(k_{\perp},k_{\parallel})$ depends on $\varepsilon(k_{\perp})$ as well.
Returning to the
general spectral form in equation (\ref{equ5b}), we see that $\varepsilon(k_{\perp})$ cancels under the
assumption of critical balance so that the dissipation is not explicitly dependent on $\varepsilon(k_{\perp})$.
However, $\overline{\omega}_r= \omega_r/k_{\parallel} v_A$ and $\overline{\gamma} = \gamma/k_{\parallel}v_A$ in (\ref{equ12}) can
be explicit functions of $k_{\parallel}$, if $\gamma(k_{\perp}, k_{\parallel})$ and $\omega_r(k_{\perp}, k_{\parallel})$ are nonlinear functions of $k_{\parallel}$.
Damping rates calculated from the \citet{lys96} approximation in (\ref{equ16}) satisfy the condition $\overline{\gamma}(k_{\perp})= \gamma(k_{\perp},k_{\parallel})/k_{\parallel} v_A$ exactly leading
to a dissipation which is independent of the energy flux and hence to the same dissipation scale for different values of the energy flux.
For normalized damping rates for KAW obtained from the hot plasma dispersion relation, 
the independence of $\overline{\gamma}$ from the parallel wavenumber cannot be
shown analytically but can be estimated numerically. 
Figure \ref{fig_kpara} shows 
the parallel wavenumber as a function of the perpendicular wavenumber
as derived in equation (\ref{k_parallel}) and (\ref{equ5c}) for four different values of $\varepsilon_0$.
The dotted line denotes the first spectral break at ion 
scales.
The 
break frequency and the original value of $\varepsilon_0=7\times10^{-16} \ \rm J m^{-3} s^{-1}$ are taken from observation 5 in \citet{ale09}.
The larger $\varepsilon_0$, the more the turbulence generates large parallel wavenumbers for the same perpendicular wavenumber.
Figure \ref{fig_gamma_epsilon} shows the hot damping rate ($\overline{\gamma}/\overline{\omega}_r$) for all 
ratios of $k_{\parallel}$ to $k_{\perp}$ from Figure \ref{fig_kpara}. All damping rates fall approximately on the same dark blue solid line. 
$\overline{\gamma}/\overline{\omega}_r$ from the \citet{lys96} approximation is shown
in the dashed line for comparison. At least for typical solar wind parameters, the normalized hot damping rates for KAW are also approximately independent of the parallel wavenumber:
$\overline{\gamma}(k_{\perp},k_{\parallel})/\overline{\omega}_r(k_{\perp}, k_{\parallel})\sim \overline{\gamma}(k_{\perp})/\overline{\omega}_r(k_{\perp})$, which leads again to the same dissipation scale for all spectra independently of the injected energy rate.\\
We can estimate this dissipation scale for solar wind turbulence similar to the HD Kolmogorov dissipation scale by equating the argument of the exponential
term in equation (\ref{E_SW}) with -1, i.e., where
the energy flux is reduced by the factor of $1/e$ and the difference is converted 
into heat or other forms of particle acceleration
\begin{linenomath*}
\begin{equation}
 1 = 2 C_K^{3/2}\int_{k_0}^{k_{\perp}} k_{\perp}'^{-1}\frac{\overline{\gamma}}{\overline{\omega}_r}(k_{\perp}') dk_{\perp}'. \label{equ13}
\end{equation}
\end{linenomath*}
Up to this scale the dissipation term is negligible or small compared to the spectral energy transport.
When we assume for mathematical simplicity the normalized damping rate to be in the form of a power law $\overline{\gamma}/\overline{\omega}_r\propto k_{\perp}^{\zeta}$,
the integral in equation (\ref{equ13}) can be solved analytically:
\begin{linenomath*}
\begin{eqnarray}
 1 = 2 C_K^{3/2} \zeta^{-1} \overline{\gamma}/\overline{\omega}_r
 \label{equ14}\\
\Rightarrow \overline{\gamma}(k_d)/\overline{\omega}_r(k_d) \sim 1.
\label{equ15}
\end{eqnarray}
\end{linenomath*}
Hence, dissipation sets in at scales $k_d = 1/l_d$ where the damping rate equals the real frequency independently of the energy cascade rate.\\
The differences of the solar wind dissipation length in comparison to the hydrodynamic dissipation length
\begin{figure*}%
\begin{center} \large{\underline{\ \ \ \ \ Hydrodynamic Turbulence \ \ \ \ \ }} \end{center}
\gridline{\fig{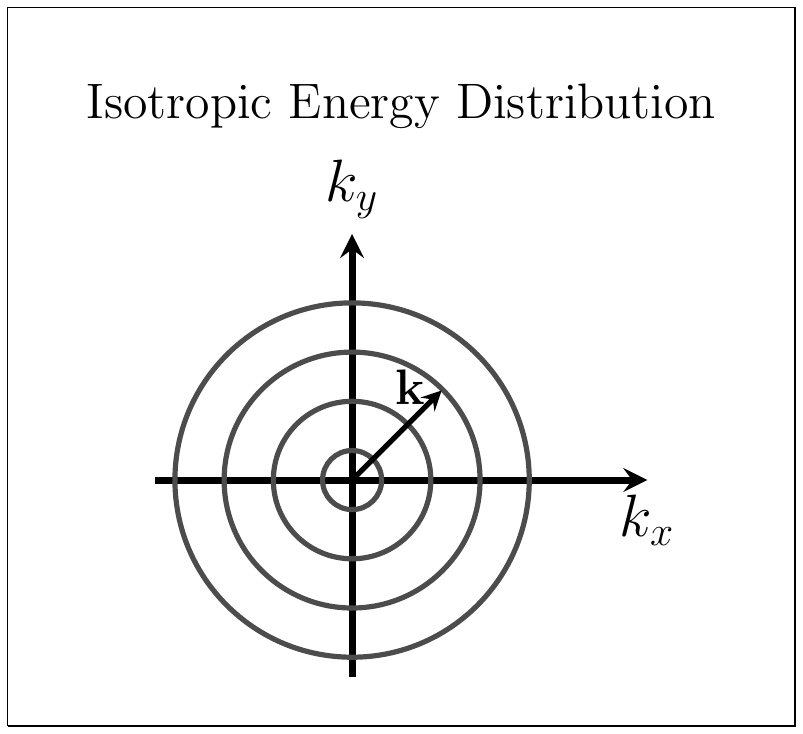}{0.36\textwidth}{\large(a)}
	  \fig{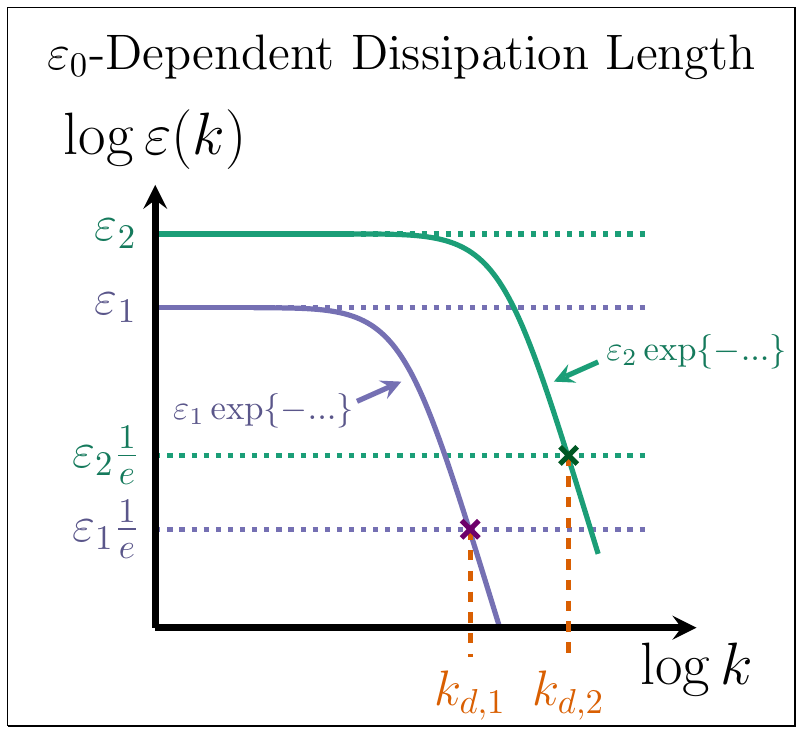}{0.36\textwidth}{\large(b)}}
      \begin{center} \large{\underline{\ \ \ \ \ Solar Wind Turbulence {\color{white}{g}}\ \ \ \ }} \end{center}
\gridline{\fig{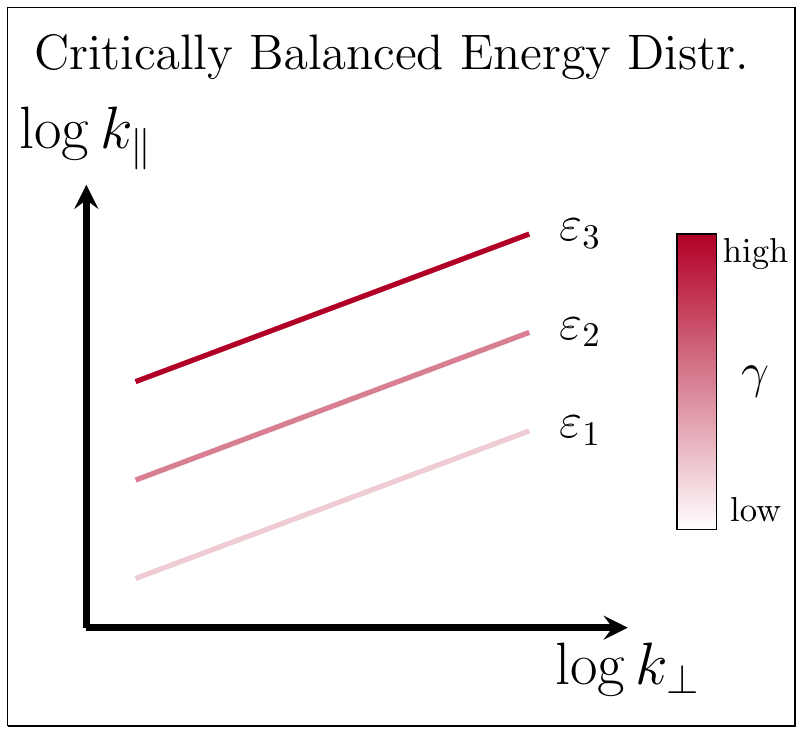}{0.36\textwidth}{\large(c)}
	  \fig{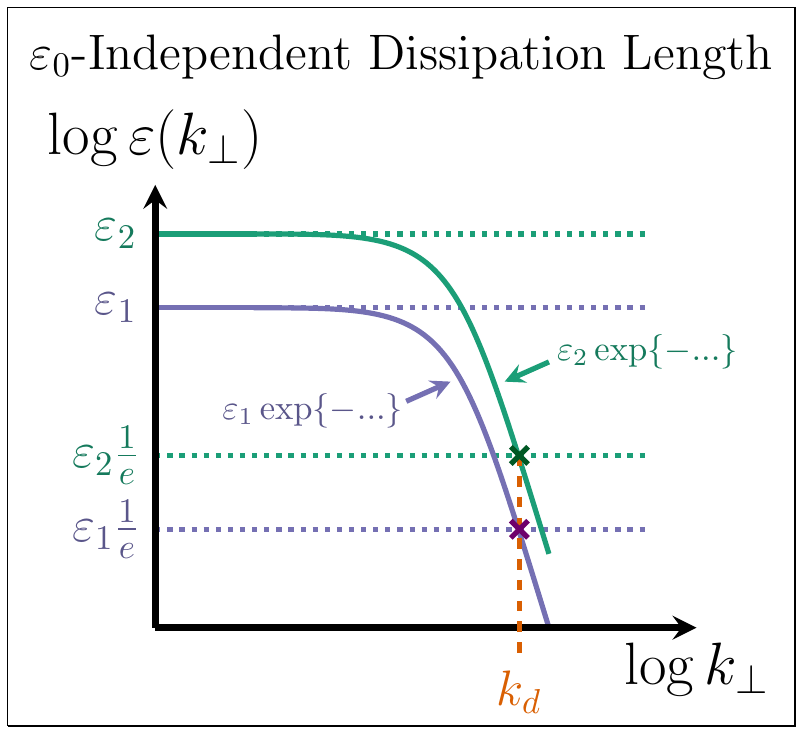}{0.36\textwidth}{\large(d)}}
\caption{
Sketch of the role of different energy cascade rates on the energy distribution in k-space (left panels) and on the energy flux $\varepsilon(k)$ (right panel) for hydrodynamic turbulence (top panels) and solar wind turbulence (bottom panels). The different values for the energy cascade rate $\varepsilon_0$ are referred to as $\varepsilon_1$, $\varepsilon_2$, $\varepsilon_3$ with $\varepsilon_1 < \varepsilon_2 < \varepsilon_3$. In panel (c), the energy distribution for solar wind turbulence is assumed to follow critical balance, which implies that larger $\varepsilon_0$ result in larger parallel wavenumbers $k_{\parallel}$. For KAW larger parallel wavenumbers additionally result in larger damping rates $\gamma$ for the same $k_{\perp}$. The larger damping rates $\gamma$ are indicated by the intensity of the red color in panel (c). The dissipation scales $k_d$ shown in orange in panel (b) and (d) are defined as the scales where the energy is reduced by a factor of $1/e$.}
\label{fig_schem1}
\end{figure*}
are sketched in Figure \ref{fig_schem1}. Top panels show the hydrodynamic case, bottom panels show
the solar wind case.
Panel (a) displays the isotropic energy distribution in HD turbulence and
panel (c) shows the anisotropic energy distribution in a magnetized plasma
under the assumption of critical balance for different values of $\varepsilon_0$ labeled $\varepsilon_3>\varepsilon_2>\varepsilon_1$.Panel (c) shows additionally in red the general intensity of
damping for different $\varepsilon_0$ for linear wave mode damping such as in our KAW model.In a critically balanced turbulence, larger values of $\varepsilon_0$ lead to larger parallel wavenumbers (see equations (\ref{equ8}) and (\ref{k_parallel})). The larger parallel wavenumbers at a given perpendicular wavenumber lead to larger damping rates. In contrast in HD turbulence, $\varepsilon_0$ has no influence on the damping rate $\gamma(k)=\nu k^2$. Following equation (\ref{equ5c}),
panels (b) and (d) illustrate schematically the influence of different values of $\varepsilon_0$ ($\varepsilon_2>\varepsilon_1$)
on the energy cascade rate $\varepsilon(k)$ and $\varepsilon(k_{\perp})$ for
hydrodynamic turbulence and solar wind turbulence, respectively. 
The dissipation length, marked by the orange dashed lines, is defined as the scale where the energy flux is reduced by a factor of $1/e$. For HD turbulence, larger $\varepsilon_0$ leads to a smaller dissipation scale, whereas the dissipation length in the solar wind plasma is independent of the energy flux.
To explain this difference in detail,
we look at the equation that describes the relative change of the energy flux (derived  from (\ref{equ5c})
\begin{linenomath*}
\begin{equation}
 \frac{1}{\varepsilon(k)}\frac{d \varepsilon(k)}{ d k} \propto - 
 \left(\frac{\varepsilon(k)}{\rho}\right)^{-1/3}\gamma(k) k^{-\kappa}.
 \label{equ_rel_change}
\end{equation}
\end{linenomath*}
For resistive HD damping the relative change of energy flux , i.e., $1/\varepsilon(k)\  d\varepsilon/dk = d/dk \ \ln \varepsilon(k)$ on the left hand side of (\ref{equ_rel_change}) depends on $\varepsilon(k)^{-1/3}$ and therefore on the energy injection rate $\varepsilon_0$. The relative change of the energy flux therefore changes depending on how strongly the turbulence is driven. Different $\varepsilon_0$ result in different amplitudes of the 
energy spectrum as well as in different exponential curves in HD turbulence.
In the case of solar wind turbulence under the assumption of a critically balanced energy distribution the situation is different.
A larger
energy flux leads to a modified anisotropic distribution of energy in k-space, i.e., larger
$k_{\parallel}$ for the same $k_{\perp}$ (see Figure 4(c)).
These larger parallel wavenumbers result in larger damping rates $\gamma \sim k_{\parallel} v_A
\overline{\gamma}(k_{\perp})\sim \varepsilon(k_{\perp})^{1/3} \overline{\gamma}(k_{\perp})$ (see colored lines and related color bar in Figure 4 (c)). By insertion of $\gamma(k)$ into (\ref{equ_rel_change}), we see that the right hand side of (\ref{equ_rel_change}) is independent of the energy flux $\varepsilon(k_{\perp})$. Therefore the relative change of the energy density and the spectral form of the energy density is independent of $\varepsilon_0$. The larger energy flux,
which drives the turbulent energy to smaller scales, is compensated by the
larger damping rates.
This compensation of a larger energy flux by larger damping rates results in the same perpendicular
dissipation scale for all values
of $\varepsilon_0$ under the assumption $\gamma(k_{\perp},k_{\parallel}) = k_{\parallel} v_A \overline{\gamma} (k_{\perp})$, which is approximately valid in the solar wind (see Figure \ref{fig_gamma_epsilon} ).\\
In addition to the analysis of dissipation length scales, our model for the spectral energy density provides the opportunity to investigate the spectral shape of the dissipation range.
There is an ongoing debate, whether the dissipation range forms an
exponential decay \citep{ale09, ale12} or follows a power law \citep{sah09,sah13}.
By looking at equation (\ref{equ_P3}), we formally see that  under the assumption of (\ref{equ_neu6}) any damping rate that is of the form $\gamma = \gamma_0(k_{\perp}/ k_0)^{\kappa-1}$
leads to a power law dissipation spectrum with a spectral index of $\kappa+4/3 C_K (\varepsilon_0/\rho)^{-1/3}\alpha_0^{-2/3} k_0^{-2/3} \gamma_0$, whereas $\gamma \propto k_{\perp}^{\kappa}$ implicates an exact exponential shape of the 
form $\exp (-l_d k_{\perp})$. Note that any 
deviation of 
$\gamma \propto k_{\perp}^{\kappa-1}$ leads to a 'quasi' exponentially shaped dissipation spectrum.
Figure \ref{fig_powerlaw} shows the damping rates, which would yield a power law (dotted line) or on the contrary an exact exponentially shaped dissipation range (dashed line) for a 
spectral index
of $\kappa=7/3$.
The KAW damping rate calculated from the hot dispersion relation for plasma parameters from observation 5 in \citet{ale09}
and for parallel wavenumbers following equation (\ref{k_parallel}) is plotted as a solid line. $\gamma_{\mbox{KAW}}$ 
follows approximately
$k_{\perp}^{2.2}$ up to the electron scales
and is thus close to the $k_{\perp}^{\kappa}$ scaling for the exponentially shaped dissipation spectrum.
At scales smaller than the electron scales, the damping rate flattens and stays approximately constant. Hence, we draw the conclusion that damping by KAWs leads to a 
'quasi' exponential decay in the
dissipation range. Further observations at sub-electron scales are necessary to see whether the flattening in the KAW damping rate has an influence on the 
magnetic spectra in this range. 
\begin{figure}
  \plotone{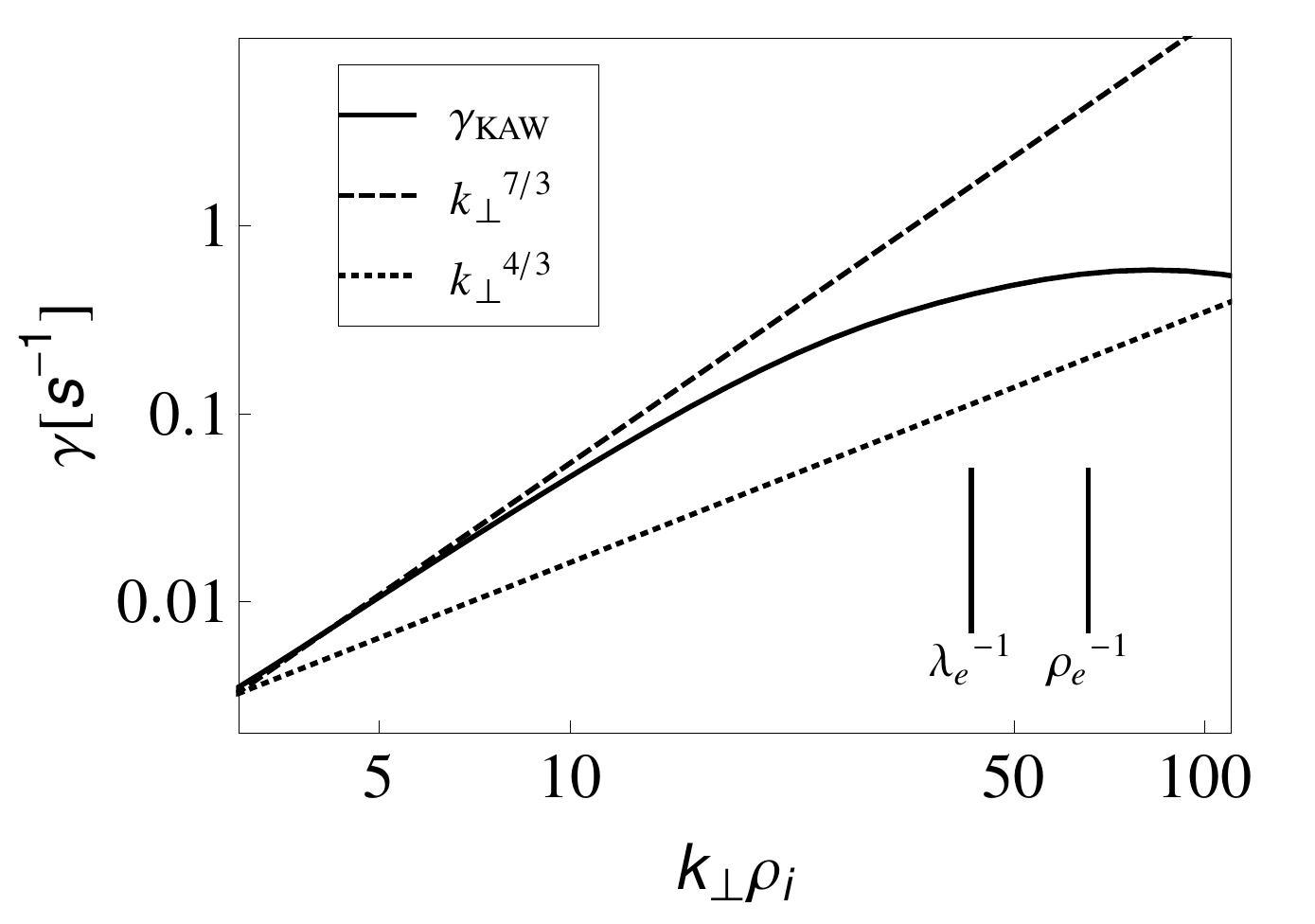}
  \caption{The solid line gives the KAW damping rate from equation (\ref{equ_det}) for the same parameters as in Figure \ref{fig_kpara}. 
  $k_{\perp}^{4/3}$ and $k_{\perp}^{7/3}$ is shown for comparison.
  $\lambda_e$ and $\rho_e$ are marked by the vertical lines.}
  \label{fig_powerlaw}
 \end{figure}

\section{APPLICATION TO THE SOLAR WIND}
\label{sec_app}
\begin{figure}
  \plotone{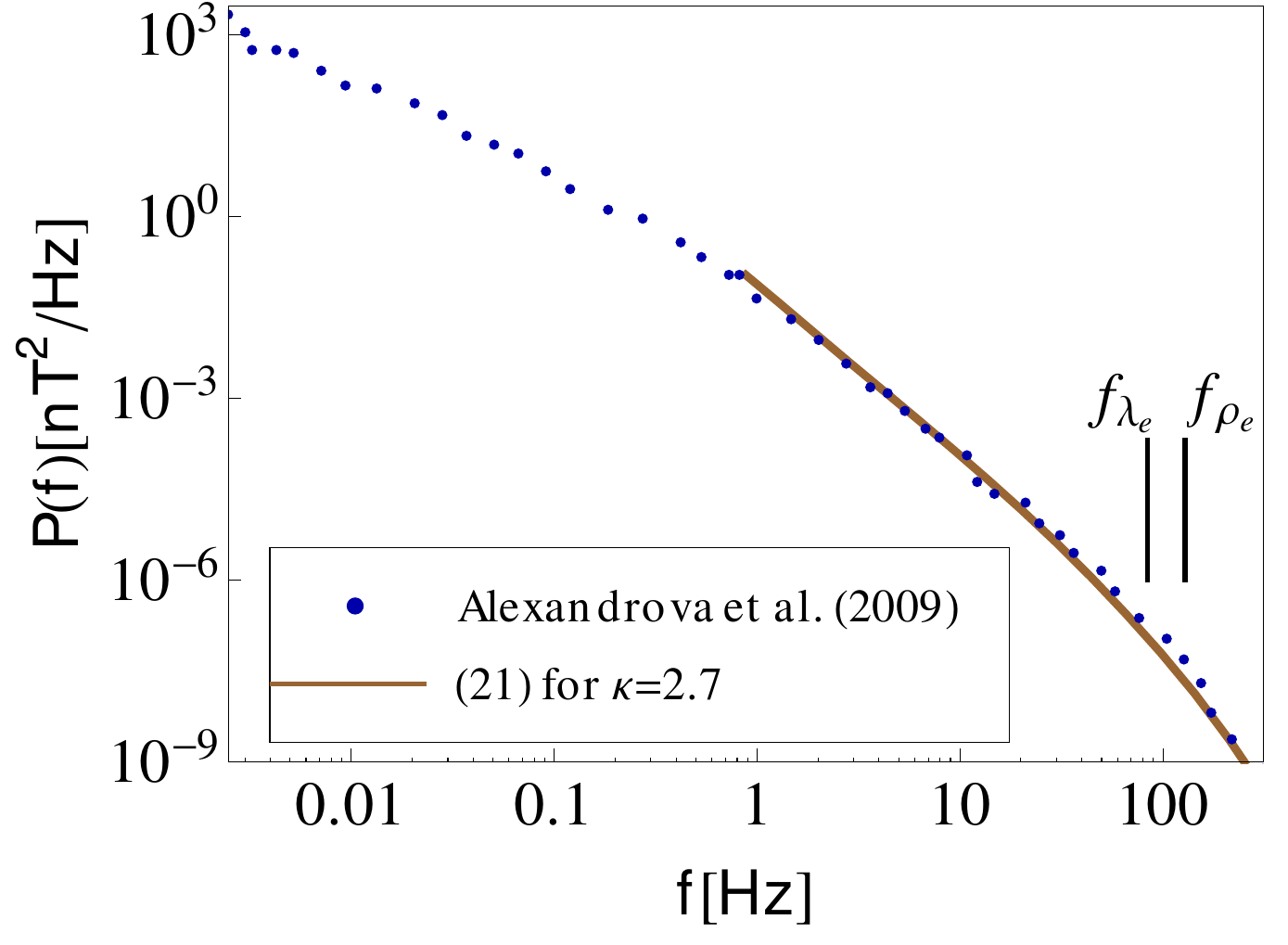}
  \caption{Equation (\ref{equ12}) for the same parameters as in Figure \ref{fig_kpara}. Observations from interval 5 in \citet{ale09} are shown in blue dots. 
  Vertical lines indicate the electron scales,
  where $f_{\lambda_e}$ corresponds to the Doppler-shifted $\lambda_e$, and $f_{\rho_e}$ to $\rho_e$.}
  \label{fig_Alex}
 \end{figure}

\begin{figure}
\gridline{\fig{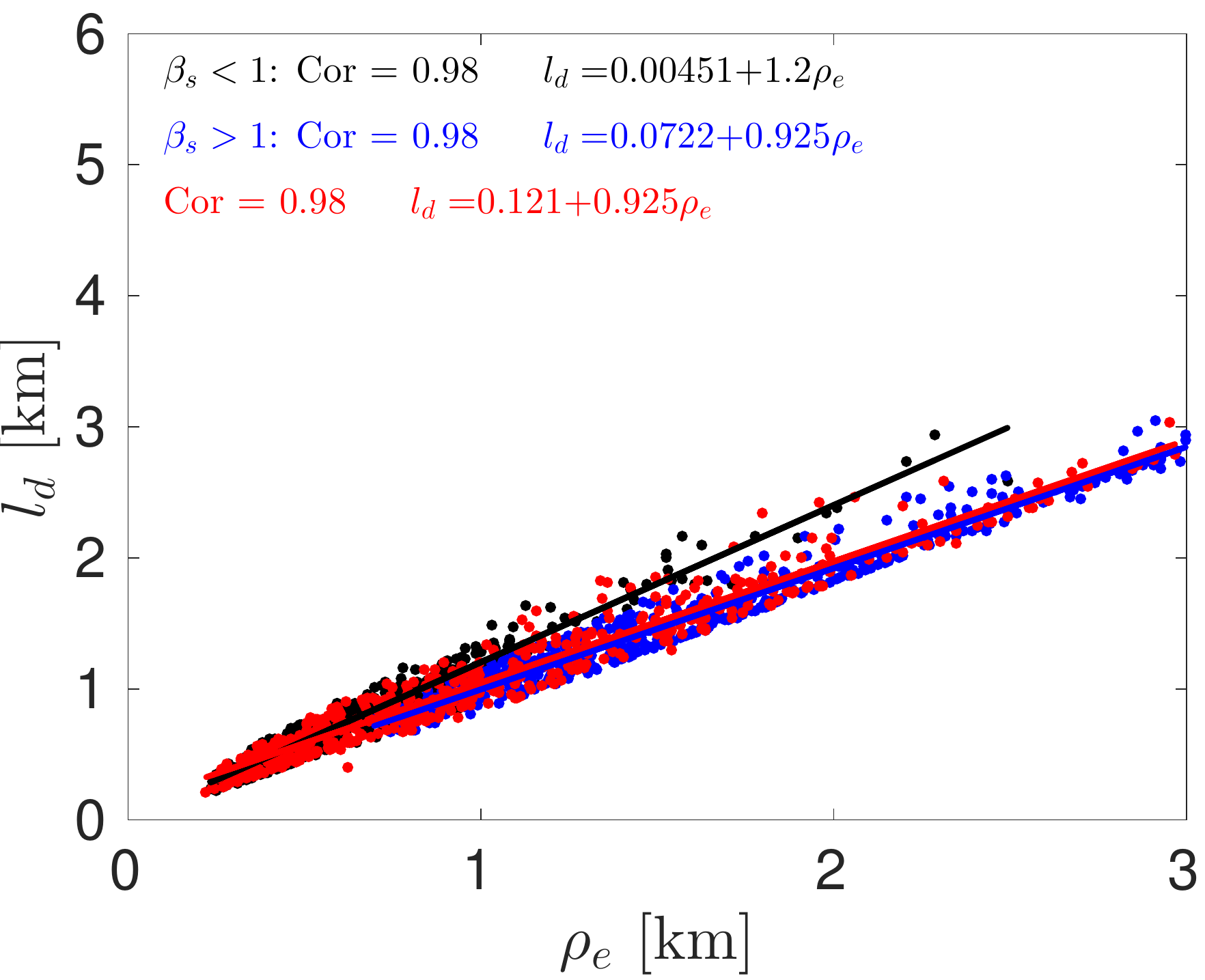}{0.5\textwidth}{(a)}}
\gridline{\fig{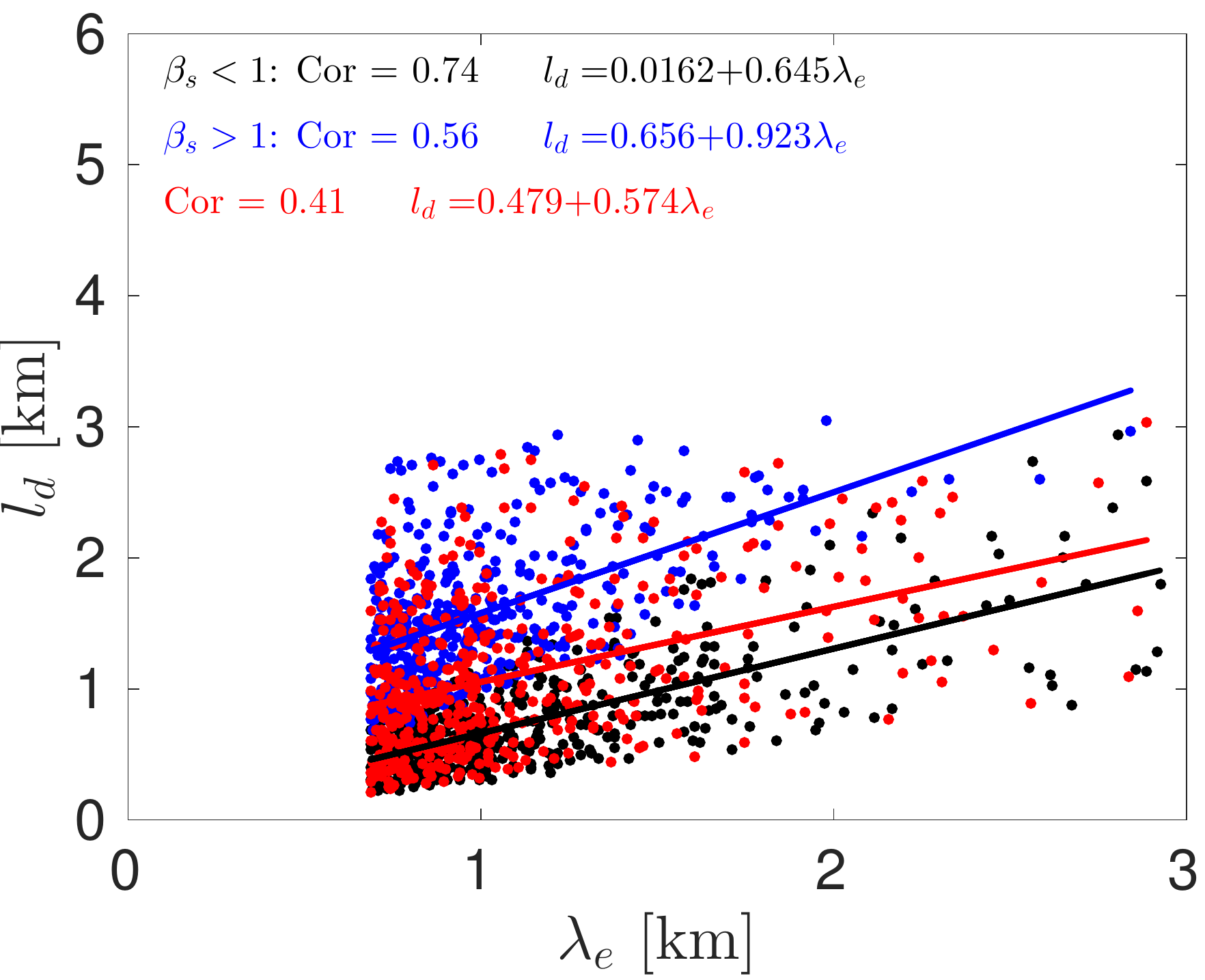}{0.5\textwidth}{(b)}}
 \caption{Results of fitting equation (\ref{equ17}) to 300 model spectra with hot damping rates. The dissipation length $l_d$ is shown as a function of the electron Larmor radius $\rho_e$ (a) and 
 of the electron inertial length $\lambda_e$ (b). The red dots show the results for $\beta_i =[0.1,10] $ and $\beta_e =[0.1,20]$, the black and blue 
 dots show separated results for small ($\beta_i, \beta_e = [0.1,1]$) and 
large plasma betas ($\beta_i= [1,10]$ and $\beta_e=[1,20]$), respectively. 
 }
  \label{fig_hot}
 \end{figure}

In this section, we quantitatively compare a model spectrum calculated with hot damping 
rates and critically balanced wavenumbers with observations in the solar wind,
followed by a statistical study to be compared with the 
statistical study of the set of observations in \citet{ale12}. The statistical study aims to 
estimate the dissipation length for varying solar 
wind conditions. Here we present the first comparison of a dissipation model with a measured magnetic spectrum at electron scales. 
The blue dots in Figure \ref{fig_Alex} show 
observed spectral energy densities by \citet{ale09} for
$B=15.5$ nT, $n=20$ cm$^{-3}$, $T_i = 61$ eV, $T_e=26$ eV, $v_S = 630$ km/s, and an angle between the mean magnetic field and the solar wind velocity  of $\Theta_{BV}=83^{\circ}$.
For low frequencies the spectrum follows $\sim f^{-1.7}$ in agreement with Kolmogorov's law and steepens on ion scales to $\sim f^{-2.8}$. 
Around the electron scales, the spectrum follows approximately an exponential function \citep{ale09}.
Our model spectrum is shown in brown for $\kappa = 2.7$ for scales below ion scales, where we have applied Taylor's hypothesis to convert wave vector
spectra into frequency spectra using $f= k_{\perp} v_S/2\pi$.
Apart from the spectral index $\kappa$,
and the Kolmogorov constant $C_K$,
our model equation has no other free parameters.
In the ranges of $\kappa=[2.2, 2.8]$ and $C_K=[1.4, 2.1]$, we find through the calculation of the root-mean-square error that the model with $\kappa=2.7$
and $C_K=1.4$ describes the data best, but combinations of $\kappa=[2.5, 2.7]$ and $C_K=[1.4, 1.8]$ lead to similar spectral densities within a root-mean-square error difference of 10\%. For the choice of 
the Kolmogorov constant, we follow \citet{bis93}. We discuss the influence of $C_K$ on energy spectra in Section \ref{sec_dis}.
Deviations from the theoretically expected value of $\kappa = 7/3\approx 2.33$ for KAW \citep{how06,sch09} may be a result of intermittency effects \citep{sal09,lio16} 
or superimposition of whistler wave fluctuations \citep{lac14}. 
Additionally, damping at electron scales results in spectral indices steeper than $7/3$ 
due to 'sampling' effects of one-dimensional spacecraft measurements \citep{pap15}.
Several different wavevectors
contribute to the spectral energy density 
at a certain spacecraft frequency, so that the sub-ion range is already affected by electron damping. For example, for a field to flow angle of $\Theta_{BV}=90^{\circ}$ this sampling effect steepens a 7/3 spectrum to 2.63 \citep{pap15}.
In order to take account of these effects,
we use 
a spectral index which fits best to the data.
The model spectrum follows in agreement with the observations a power law 
at the large scales and forms a 'quasi' exponential decay at the electron scales. 
Hence, the observed exponential form
of the dissipation range in
the observations seems to be compatible with electron Landau damping of kinetic 
Alfv\'{e}n waves at least for this set of observations. \\
For further insight into the spectral behavior for varying parameters,
we perform a statistical study with our model similar to the statistical study 
of 100 observed spectra by \citet{ale12}.
They fit an exponential function
with a characteristic dissipation scale $l_d$  and  with a power law pre-factor 
\begin{linenomath*}
\begin{equation} P_A(k_{\perp}) = A k_{\perp}^{-\alpha} \exp (-k_{\perp} l_d) \label{equ17}\end{equation}
\end{linenomath*}
to the solar wind spectra.
The study by \citet{ale12} finds that the variations of $l_d$ due to different solar wind conditions are related to the variations of the electron Larmor radius, $l_d \sim 1.35 \rho_e$, 
with a high correlation coefficient of 0.7. The correlation between $l_d$
and the electron inertial length $\lambda_e$ is much weaker with a correlation coefficient of 
0.34. The authors assume that 
the dissipation range in the analyzed set of spectra follows
a universal structure of the form of equation
(\ref{equ17}) for all solar wind parameters. Here we use the same parameter ranges
as the observed ones for the magnetic fields, the temperature
ratios and the
number densities:
$B \in [2,20]$ nT, $T_i/T_e \in [0.5,5]$ and $n_i=n_e \in [3,60]$ cm$^{-3}$. The results of fitting equation (\ref{equ17}) 
to our model through a a least mean square fit are shown in
Figures \ref{fig_hot} (a) and \ref{fig_hot} (b).
The red dots show the results for a wide range of ion and electron plasma betas
($\beta_i \in [0.1,10] $ and $\beta_e \in [0.1,20]$),
the black and blue dots show separated results for small ($\beta_i, \beta_e \in [0.1,1]$) and 
large plasma betas ($\beta_i \in [1,10]$ and $\beta_e\in[1,20]$), respectively.
For every model spectrum, the parameters are chosen randomly within the given parameter 
ranges using logarithmic distributed values for the temperature ratio and the plasma beta and 
linear distributed values for the others.
We find a very high correlation for the electron Larmor radius of 0.98 and a dissipation length $l_d \sim 0.9 \rho_e$, which is similar to the observed value by \citet{ale12}.
Also in agreement with the observational study by \citet{ale12}, Figure \ref{fig_hot} (b)
shows a much weaker correlation of 0.41 between the dissipation length $l_d$ 
and the electron inertial length $\lambda_e$.
This correlation is mainly due to intervals, where $\beta_e \approx 1$, 
which means that the inertial length is comparable to the Larmor radius. 
Another possible explanation is, that also the inertial length is related to the
dissipation scale for some solar wind conditions.
For example for small electron plasma betas and low
temperatures the electron Larmor radius is very small. 
In this case, the
turbulence might dissipate on an alternative scale, e.g., the electron 
inertial length, which is reached first by the turbulent cascade. 
In order to look into this hypothesis, we
study the dissipation length separately for small (black line) and large
plasma betas (blue line).
Indeed, the correlation between the dissipation length and the electron inertial 
length is higher for small plasma 
betas with a correlation coefficient of $0.74$ than for large 
plasma betas with a correlation coefficient of $0.56$. 
Additionally, the estimated dissipation length in the case of small plasma betas
($l_d \sim 1.2 \rho_e$) is slightly 
larger than in the large beta case ($l_d \sim 0.9 \rho_e$) suggesting that the energy is dissipated at scales larger than the electron gyroradius.

\section{DISCUSSION}
\label{sec_dis}
Here we discuss a number of assumptions that have been made in the construction of 
our solar wind dissipation model at electron scales.
A range of the Kolmogorov constant $C_K = [1.4, 2]$ in the solar wind was determined from experimental data and nonlinear simulations \citep{bis93}. 
In this study the constant is taken to be $C_K = 1.4$ in both the MHD and kinetic regime. 
However, the 'constant' may depend on the plasma parameters. For higher $C_K$ the argument of the exponential term
in equation (\ref{equ12}) is larger and therefore the effect of damping is increased in comparison to the nonlinear energy transfer.
This variation of the Kolmogorov constant leads to an uncertainty in the magnetic spectra, but without any influence on the general physical description.\\
Here we use critical balance to obtain the anisotropy of the cascade of energy to smaller scales. This assumption is valid only for strong turbulence. On the contrary, there is no parallel energy cascade
in weak turbulence \citep{gol94}. However, with increasing $k_{\perp}$  the nonlinear interactions become so strong that the assumption of weakness is no longer valid.
Therefore the turbulence is either already strong from the beginning or will eventually become strong for increasing $k_{\perp}$. Yet, our model is not able to handle a changing from 
strong to weak turbulence when the collisionless damping reduces the amplitudes of the nonlinear interactions to a limit, where weak turbulence should be applied (See \citet{how11} for 
a weakened cascade model).\\
The dissipation model presented here is similar to two earlier models, which also contain a nonlinear energy cascade and collisionless damping. \citet{pod10} computed numerically 
the damping rate from the hot plasma dispersion relation. They conclude that a KAW energy cascade is almost completely dissipated before reaching the electron scales
due to strong Landau damping. This would imply that the energy cascade to the electron scales must be supported by wave modes other than the KAW.
\citet{how11} argued, that they underestimated the weight of the nonlinear energy cascade in comparison to the dissipation (here described by $C_K$), leading to overestimated damping rates.
The cascade model in \citet{how08} employs the damping rates obtained from gyrokinetic theory. The authors find in agreement to our results an exponential shaped dissipation range
for moderate damping with $\beta_i=1$ for $T_i/T_e=1$. For strong damping($\beta_i=0.01$ and $T_i/T_e=1$) the spectra show sharp cut offs. In \citet{how11} it is assumed that in a model with only local interactions
the damping dominates over the energy transfer in the case of strong damping. Therefore they constructed a weakened cascade model with nonlocal interactions. Following \citet{sch09}, damping can be considered strong if the decay time $1/\gamma$ is shorter or comparable to the wave period $2 \pi /\omega_r$. Figure \ref{fig_LL96T} shows that damping at $k_{\perp}\rho_i=1$ is relatively weak for typical solar wind parameters ($\beta_i \gtrsim 1$, $T_i/T_e \approx [0.5, 5]$),
thus the nonlocal effects should play a minor role in interpreting the observed energy spectra.\\
Our dissipation model is a linear model in the sense that it linearly combines the non-linear cascade towards smaller length scales and a process transferring magnetic field energy to particle energy. The mutual feedback of these processes might become stronger at small scales, where the dissipation rates become strong.
However, we expect our model to still capture important aspects of the physics around electron scales. In our model we neglect physics on scales significantly beyond the electron scales, e.g., a possible third electrostatic turbulent cascade \citep{sch16}.\\
For mathematical simplicity, we solve the hot plasma dispersion relation assuming Maxwellian distributions of 
protons and electrons with no temperature anisotropies.
Observations of particle distributions show deviations from a Maxwellian due to the
weakly collisional nature of 
the solar wind \citep{hun70, fel73, goo76}. Measured electron distribution
functions are composed of an almost 
Maxwellian and isotropic core for electrons with energy below 50 eV and a highly
anisotropic
halo representing electrons
of higher energy \citep{bri09}.
Likewise, observations of proton distribution functions indicate anisotropies between
the temperatures parallel and perpendicular to the magnetic field and bump-like 
deformations at high energy \citep{mar82}.
However, due to instabilities limiting the scope of the deformations, 
the measured deformation of the thermal distribution function is not 
as strong as expected \citep{bri09}.\\

\section{CONCLUSIONS}
\label{sec_con}
We present an analytic dissipation model to describe turbulence at electron scales. It combines the energy transport from large to
small scales and the dissipation by collisionless damping of KAWs. 
The model provides the possibility to analyze and interpret observations of turbulent fluctuations in 
the dissipation range with in principle arbitrary spectral index in the electron inertial range. The key
results of our study are: 
A direct comparison of our model with observed spectra by \citet{ale09,ale12} in the solar wind shows that 
damping by kinetic Alfv\'{e}n waves can explain the 'quasi' exponential spectral structure of the
dissipation range at least for the observed solar wind conditions. 
The dissipation model provides an explanation for the independence of 
the dissipation scale from 
the energy cascade rate,
which is a remarkable difference compared to hydrodynamic
turbulence. This difference is due
to the anisotropic nature of the plasma turbulence, i.e., due to a combination of
critically balanced turbulence and a dispersion relation proportional to the parallel wavenumber.
The critical balance 
assumption influences the energy cascade in a way, that the more energy is injected
at the driving scales, the more effective the damping rate gets.
A statistical study
of model spectra confirms the high correlation between the
dissipation length and the electron Larmor radius, as was
reported in \citet{ale12}. Therefore
the Larmor radius may 
play the role of a dissipation scale in solar wind turbulence.
Our dissipation model can easily be
applied to other turbulent systems, e.g., planetary magnetospheres for the prediction of spectral energy densities.
\acknowledgments
We thank A.~A. Schekochihin, G.~G. Howes, O. Alexandrova and M. von Papen for helpful discussions.
\appendix
\section{Hot Plasma Dispersion Relation}
\label{app_hot}
\subsection{Dielectric Tensor}
\label{app_tensor}
For a nonrelativistic plasma with Maxwellian distributed electrons and protons with no zero-order drift velocities, the elements of the
dielectric tensor can be cast in the form \citep[e.g.,][]{che84,sti92}\\
\begin{linenomath*}
\begin{eqnarray}
 \epsilon_{xx} &=& 1+ \sum_s\frac{\omega_{ps}^2}{\omega^2}\xi_{0s} \sum_{n=-\infty}^{\infty} n^2 \frac{\Gamma_n(\mu_s)}{\mu_s} Z(\xi_{ns})\\
 \epsilon_{yy} &=& 1+ \sum_s\frac{\omega_{ps}^2}{\omega^2}\xi_{0s} \sum_{n=-\infty}^{\infty} \left\{n^2 \frac{\Gamma_n(\mu_s)}{\mu_s} - 2 \mu_s 
 \Gamma_n'(\mu_s) \right\}
 Z(\xi_{ns})\\
 \epsilon_{zz} &=&1 - \sum_s\frac{\omega_{ps}^2}{\omega^2}\xi_{0s} \sum_{n=-\infty}^{\infty} \xi_{ns} \Gamma_n(\mu_s) Z'(\xi_{ns})~~\\
 \epsilon_{xy} &=&   i \sum_s\frac{\omega_{ps}^2}{\omega^2}\xi_{0s} \sum_{n=-\infty}^{\infty} \Gamma_n'(\mu_s) n Z(\xi_{ns})\\
 \epsilon_{xz}&=& - \sum_s\sgn(q_s)\frac{\omega_{ps}^2}{\omega^2}\xi_{0s} \sum_{n=-\infty}^{\infty}  \frac{1}{\sqrt{2\mu_s}} n \Gamma_n(\mu_s) Z'(\xi_{ns})\\
 \epsilon_{yz}&=&  i \sum_s\sgn(q_s)\frac{\omega_{ps}^2}{\omega^2}\xi_{0s} \sum_{n=-\infty}^{\infty} \Gamma_n'(\mu_s) Z'(\xi_{ns})\sqrt{\frac{\mu_s}{2}},
\end{eqnarray}
\end{linenomath*}
where $\omega_{ps}= (n_s q_s^2/\epsilon_0 m_s)^{1/2}$ is
the plasma frequency of species $s$ (with $n_s$ the number density, $q_s$ the charge,
and $m_s$ the particle mass), $\Omega_s=q_s B/m_s$ is the gyrofrequency of species $s$ (negative
for electrons), $\xi_{ns}= (\omega-n\Omega_s)/k_{\parallel} v_s$, $v_s=(2 k_B T_s/m_s )^{1/2}$ is the 
thermal speed of species $s$ (with the Boltzmann constant $k_B$ and the temperature $T_s$), and $\mu_s = 0.5 k_{\perp}^2 \rho_s^2$ (with the Larmor radius
$\rho_s = v_s/\Omega_s$).
The function $Z(\xi)$ is the plasma dispersion function, which was introduced by \citet{fri61}. Its 
derivative is given by $Z'(\xi) = -2 - 2 \xi Z(\xi)$. $\Gamma_n(\mu_s) = e^{-\mu_s} I_n(\mu_s)$, where
$I_n§$ is the modified Bessel function of the first kind of order $n$. Note that the derivative of $\Gamma_n$ is given 
by $\Gamma_n'(\mu_s)= \left( I_n'(\mu_s)-I_n(\mu_s)\right) e^{-\mu_s}$.
\subsection{Numerical Implementation}
\label{app_num}
If we make no assumptions for the wave frequency and the plasma beta, the full system described by
equation (\ref{equ_det}) needs to be solved numerically to find the wave frequency for given plasma parameters.
 In contrast to most previous studies, we do not apply the eight-pole approximation (Pad\'{e} approximation) to evaluate the
 plasma dispersion function $Z(\xi)$ \citep[see e.g.,][]{rön82} but evaluate the $Z$ function directly 
 in the form
 \begin{linenomath*}
 \begin{equation}
  Z(\xi) = i \sqrt{\pi} \exp(-\xi^2) ~\mathrm{erfc}(\xi)
 \end{equation}
 \end{linenomath*}
with the complementary error function erfc$(\xi)$ \citep[e.g.,][]{abr64}. In this way, we make sure that the damping rates 
are evaluated correctly even for heavily damped waves, i.e., $\operatorname{Im}(\omega)> -k_{\parallel}$ or
$ \operatorname{Im}(\omega) > - \operatorname{Re}(\omega-n\Omega)$ \citep{rön82}.\\
A two-dimensional Newton's method root search in the complex frequency plane is used here to find
the solution of equation (\ref{equ_det}). To ensure accurate results for high perpendicular 
wavenumbers, the number of sum elements that are kept is about the same as $k_{\perp} \rho_i$ 
\citep{how06}.
We implemented an iterative root search to track the required wave mode from small wavenumbers to large wavenumbers. 
An initial guess of the frequency is set at a given initial wavenumber (e.g., MHD Alfv\'{e}n wave frequency to track the kinetic Alfv\'{e}n wave).
At neighboring wavenumbers, the solution is then found by using the previously obtained frequency as an initial guess.

\listofchanges

\end{document}